\documentclass[12pt]{article} 

\usepackage{graphicx}
\usepackage{amssymb}
\usepackage{amsmath,amsfonts}
\usepackage{slashed}
\usepackage{mathtools}
\usepackage{enumerate}
\usepackage{authblk}

\begin{document}

\normalsize

\begin{center}
\Large \bf Equilibrium of Charges and Differential Equations Solved by Polynomials II
\end{center}

\vspace{2mm}

\begin{center}
Igor Loutsenko\footnote{Corresponding author: igor.loutsenko@gmail.com, loutseni@crm.umontreal.ca} and Oksana Yermolayeva
\end{center}

%

\begin{center}
Laboratoire de Physique Math\'ematique,
CRM, Universit\'e de Montr\'eal
\end{center}

\vspace{5mm}

\begin{abstract}

{

We continue study of equilibrium of two species of 2d coulomb charges (or point vortices in 2d ideal fluid) started in \cite{Lv}. Although for two species of vortices with circulation ratio $-1$ the relationship between the equilibria and the factorization/Darboux transformation of the Schrodinger operator was established a long ago, the question about similar relationship for the ratio $-2$ remained unanswered. Here we present the answer: One has to consider Darboux-type transformations of third order differential operators rather than second order Schrodinger operators. Furthermore, we show that such transformations can also generate equilibrium configurations where an additional charge of a third specie is present. Relationship with integrable hierarchies is briefly discussed.
}

\end{abstract}

\section{Introduction}
\label{vortex}

Systems of point vortices are weak solutions of the Euler equation for a flow of two-dimensional ideal fluid. Steady irrotational flows, which are singular only at a finite number of points are of a particular interest. When such a flow occupies the whole plane and its velocity vanishes at infinity, positions of singularities, i.e. ``positions of vortices", $z_i=x_i+{\rm i}y_i$ are roots of the system of $N$ algebraic equations
\begin{equation}
\sum_{j=1, j\not=i}^N \frac{Q_j}{z_i-z_j}=0, \quad i=1, \dots, N ,
\label{equilibrium}
\end{equation}
where $N$ is the number of vortices and $Q_i$ is a value of circulation of the $i$th vortex. Derivation of (\ref{equilibrium}) from the Euler equation can be found e.g. in \cite{LY}. For a general introduction to vortex dynamics and equilibrium, see e.g. \cite{Ar1, Ar2, Cl, KWCC, LY, ON1, ON2} and references therein.

System (\ref{equilibrium}) has another interpretation: It describes equilibrium of $N$ charges interacting through 2d coulomb potential. Solutions of (\ref{equilibrium}) are stationary points of the coulomb energy function $E=-\sum_{i<j}Q_iQ_j\log|z_i-z_j|$. In this interpretation $Q_i$ are values of the coulomb charges.

System (\ref{equilibrium}) is invariant under rigid motions and scaling of the plane. From invariance of the stationary value of the coulomb energy under scaling of the plane it follows that the system has solutions only if $\sum_{i<j}Q_iQ_j=0$. O'Neil showed \cite{ON1} that for almost all values of $Q_i$, satisfying the above condition, there are exactly $(N-2)!$ distinct configurations, i.e. distinct, modulo translations, rotations and scaling of the plane, solutions of (\ref{equilibrium}). However, there are ``resonant" values of $Q_i$, where number of configurations is infinite.

In this paper we mainly deal with two families of resonant cases\footnote{For more details on resonant cases see e.g. \cite{ON2, VD} and references therein.} i) two species of vortices with the circulation ratio -1,  and ii) two species with the circulation ratio -2. These families are infinite sequences of configurations depending on increasing number of continuous free parameters. They are related to rational solutions of hierarchies of integrable PDEs. Families with an additional vortex of the third specie are considered in Section \ref{Even}.

To set the problem, we first need to outline history of the question and introduce notations.

Without loss of generality, for a system consisting of two distinct species of $N=l+m$ vortices with circulation ratio $-\Lambda$, we can choose
\begin{equation}
Q_i=\left\{
\begin{array}{ll}
-1, & i=1\dots l\\
\Lambda, & i=l+1 \dots l+m
\end{array}
\right. .
\label{Lambda}
\end{equation}
For such a system the equilibrium condition (\ref{equilibrium}) can be rewritten in terms of bilinear differential equation for two polynomials $p(z)$ and $q(z)$ (of the $l$th and $m$th degree correspondingly)
\begin{equation}
p''q-2\Lambda p'q'+\Lambda^2pq''=0,
\label{bilinear}
\end{equation}
where prime denotes differentiation wrt $z$ and polynomials
\begin{equation}
p(z)=\prod_{i=1}^l (z-z_i), \quad q(z)=\prod_{i=1}^m (z-z_{l+i}) .
\label{pzqz}
\end{equation}
Since we consider two distinct species, $p$ and $q$ do not have common or multiple roots (for cases with common/multiple roots see e.g. \cite{DK2} and references therein).

\begin{figure}
\centering
\includegraphics[width=155mm]{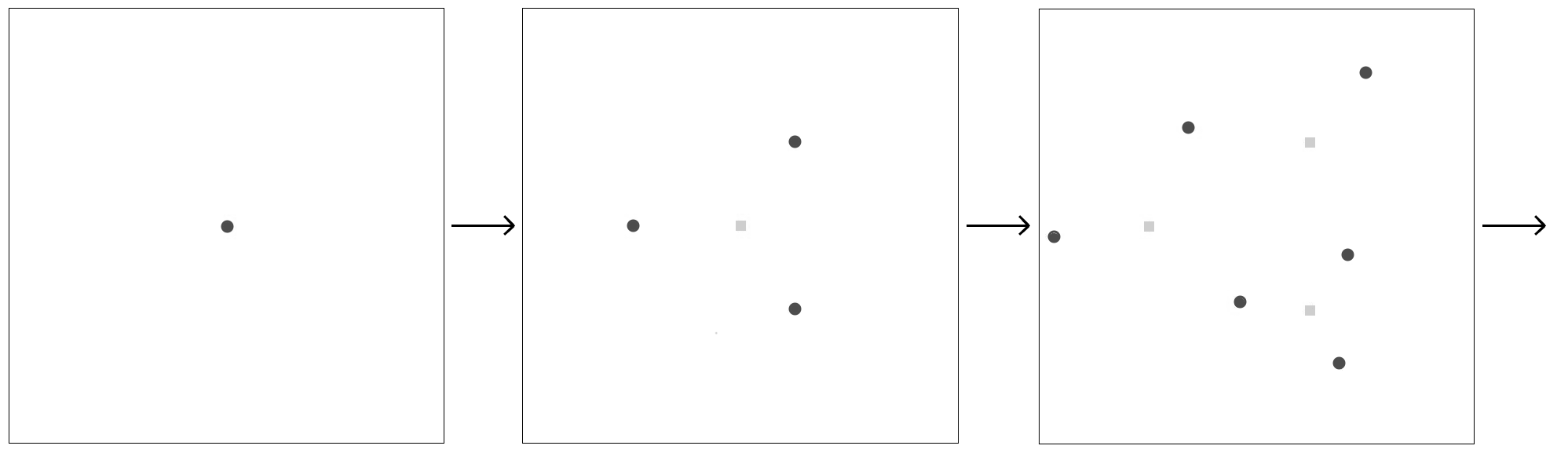}
\label{AMRoots}
\caption{$\Lambda=1$ case: Configurations corresponding to subsequent pairs $(P_0,P_1)$, $(P_1,P_2)$ and $(P_2,P_3)$ of the Adler-Moser polynomials. Roots of the first polynomial in the pair (positions of charges -1) are shown by the gray squares, while roots of the second polynomial in the pair (positions of charges +1) are shown by black circles. In this example $s_1=1, s_2=1+5{\rm i}$.}
\end{figure}

In the first resonant case circulation ratio equals -1, $\Lambda=1$ (two species with opposite circulations of equal magnitude, see (\ref{Lambda})) and Eq. (\ref{bilinear}) becomes
\begin{equation}
p''q-2 p'q'+pq''=0.
\label{Tkachenko}
\end{equation}
It first appeared and was completely solved in the paper by Burchnall and Chaundy \cite{BC}. Later this equation emerged in work by Tkachenko \cite{Tk} in context of vortex equilibrium (``Tkachenko equation"). Bartmann \cite{Bar} identified bilinear equation (\ref{Tkachenko}) with recurrence relation for the Adler-Moser polynomials $P_n(z)$
$$
q(z)=P_n(z), \quad p(z)=P_{n+1}(z), \quad n \in \mathbb{Z}_{\ge 0} ,
$$
which are polynomial $\tau$-functions of the KdV hierarchy \cite{AM}. The $n$th Adler-Moser polynomial is of degree $n(n+1)/2$ in $z$ and also depends non-trivially on additional $n-1$ free parameters $s_i$, i.e.
$$
\begin{array}{l}
P_0=1, \\
P_1=z, \\
P_2=z^3+s_1,\\
P_3=z^6+5s_1z^3+s_2z-5s_1^2,\\
\dots \\
P_n=P_n(z, s_1, \dots, s_{n-1}),\\
\dots
\end{array}
$$
Examples of several first configurations of this family are shown on Figure \ref{AMRoots}.

The Adler-Moser polynomials also satisfy the first-order differential recurrence relation \cite{AM, BC}
\begin{equation}
P'_{n+1}P_{n-1}-P'_{n-1}P_{n+1}=(2n+1)P^2_n .
\label{AM3}
\end{equation}
The rational solutions of the KdV hierarchy are proportional to the second logarithmic derivative of the Adler-Moser polynomials \cite{AM}
$$
u_n=-2(\log P_n)'' .
$$
They are potentials of the Schrodinger operators (in what follows $\partial$ stands for differentiation wrt $z$)
$$
H_n=-\partial^2+u_n,
$$
that are the ``zero energy level" Darboux transforms of the free Schrodinger operator $H_0=-\partial^2$. Operator $H_n$ is the (rational) Lax operator of the KdV hierarchy.

The Darboux transformation method allows to represent solutions of (\ref{Tkachenko}) in the Wronskian determinant form \cite{AM, BC}. It also allows to construct overcomplete rings of commuting differential operators that contain $H_n$ \cite{BC, BC1}.

In his work \cite{Bar}, Bartmann also considered the second resonant case (circulation ratio -2). According to (\ref{Lambda}), it corresponds to specification $\Lambda=2$ of Eq. (\ref{bilinear})
\begin{equation}
p''q-4 p'q'+4pq''=0.
\label{L2}
\end{equation}
This case was further studied in \cite{DK1, DK2, KWCC, Lv, ONCS}, in particular in \cite{Lv}, in relation with a generalization of the problem initially posed by Burchnall and Chaundy \cite{BC}. In this generalization one asks the question: When two primitives $\int\frac{p^{2/\Lambda}}{q^2} dz$ and $\int\frac{q^{2\Lambda}}{p^2} dz$, where $p$ and $q$ are polynomials that do not have common or multiple roots, are both rational functions of $z$? Solution to this problem is equivalent to solution of the bilinear equation (\ref{bilinear}) and exists when $\Lambda\in\{1, 1/2, 2\}$ (cases $\Lambda=1/2$ and $\Lambda=2$ are equivalent). As a consequence, equation (\ref{L2}) can be solved  completely \cite{Lv, LY} using an approach similar to that by Burchnall and Chaundy \cite{BC}. Similarly to (\ref{AM3}), in the $\Lambda=2$ case, pairs $q_n,p_n$ or $q_n, p_{n-1}$ solving (\ref{L2}) also satisfy the first-order recurrence relation \cite{Bar, Lv}
\begin{equation}
\begin{array}{ll}
q'_{n+1}q_n-q_{n+1}q'_n=(3n+1)p_n, \\
p'_np_{n-1}-p_np'_{n-1}=(6n-1)q_n^4
\end{array} , \quad n \in \mathbb{Z} .
\label{Bartman}
\end{equation}
However, in difference from the $\Lambda=1$ case, the procedure for constructing solutions by applications of Darboux transformations remained unknown.

Krishnamurthy et al. \cite{KWCC} found an integral transformation that produces all the above mentioned resonant families (among other examples). When $\Lambda=1$, it reduces to the Darboux transformation. When $\Lambda=2$, it is a composition of the Darboux transformation and exponentiation (see \cite{LY} for details), but not a Darboux-type transformation for a differential equation.

Demina and Kudryashov observed that solutions of (\ref{L2}) are related to the Sawada-Kotera hierarchy \cite{DK1}. This led us to the idea to consider the Darboux transformation for differential operators of the third order, since the Lax operator of the Sawada-Kotera hierarchy has this order \footnote{Demina and Kudryahov considered different aspect of the relationship, namely connection with self-similar reductions of Sawada-
Kotera and Kaup-Kupershmidt hierarchies studied earlier by Kudryashov \cite{Ku}. }. The Darboux transformations for the third-order operators were studied in \cite{ABO, AtNi, Ni}. To generate solutions of (\ref{L2}), we will introduce ``zero level" extension of such transformations. Additionally, we will show that apart from solutions of (\ref{L2}), configurations with an additional charge of the third specie (previously found by O'Neil and Cox-Steib in \cite{ONCS}) can be also constructed with help of introduced transformations.

Despite similarities between the $\Lambda=1$ and $\Lambda=2$ cases, the fact that the $\Lambda=2$ case is related to Darboux transformations of differential operators of the third order, rather than those of the second order Schrodinger operators, implies some qualitative difference with the $\Lambda=1$ case from the vortex equilibrium viewpoint. In particular, translating configurations, which are solutions of a generalization of (\ref{equilibrium}), cannot be now constructed by the Darboux transformation method.

Another unlikeness between the cases arises when one looks for determinant representations of configurations. These and other dissimilarities are discussed in the conclusion section. There, we also briefly go into relationships between the $\Lambda=2$ configurations and rational solutions of the Sawada-Kotera and Kaup-Kupershmidt hierarchies in context of Pfaffian/determinant representations of these configurations.

Before going into detail on the third-order operators, we will recall method of generation of equilibrium configurations through Darboux transformations of the second-order Schrodinger operators (for more detail and review, see e.g. \cite{LY}).

\section{Darboux Transformations of Second Order Operators and Equilibrium of Charges}
\label{DT2}

The Schrodinger operator
\begin{equation}
H=-\partial^2+u
\label{Schrodinger}
\end{equation}
can be written as
\begin{equation}
H=A^*A+\eta,
\label{HAA}
\end{equation}
where $A$ and $A^*$ are (formally adjoint) first order differential operators
\begin{equation}
A^*=-\varkappa^{-1}\partial\varkappa=-\partial-\varkappa'/\varkappa, \quad A=\varkappa\partial\varkappa^{-1}=\partial-\varkappa'/\varkappa.
\label{AstarA}
\end{equation}
Here, the function $\varkappa=\varkappa(z)$ is an eigenfunction of $H$ corresponding to eigenvalue $\eta$
\begin{equation}
H\varkappa=\eta\varkappa .
\label{Heta}
\end{equation}
Permuting factors $A^*$ and $A$ in (\ref{HAA}), we obtain the new Schrodinger operator $\hat H$ :
\begin{equation}
\hat H=AA^*+\eta=-\partial^2+\hat u, \quad \hat u=u-2(\log\varkappa)'' .
\label{DTH}
\end{equation}
From (\ref{HAA}) and (\ref{DTH}) it follows that for any constant $\lambda$
$$
A(H-\lambda)=(\hat H-\lambda)A .
$$
Acting by both sides of the above identity on the eigenfunction $\psi=\psi(z;\lambda)$ of $H$: $H\psi=\lambda\psi$, we come to the conclusion that the transformation
\begin{equation}
\hat \psi=A \psi=\psi'-\frac{\varkappa'}{\varkappa}\psi 
\label{DTpsi}
\end{equation}
maps eigenfunctions corresponding to eigenvalue $\lambda$ of $H$ to those of $\hat H$. Thus, given the seed eigenfunction $\psi(z;\eta)=\varkappa$ corresponding to the seed eigenvalue $\eta$ of $H$ (see (\ref{Heta})), the new operator $\hat H$ and its eigenfunctions $\{ \hat\psi=\hat \psi(z;\lambda)$ : $\hat H\hat\psi=\lambda\hat\psi \}$ can be obtained from  $H$ and its eigenfunctions $\{ \psi=\psi(z;\lambda)$ : $H\psi=\lambda\psi \}$.
The transformation $H\to\hat H$, $\psi\to\hat\psi$ given by (\ref{DTH}, \ref{DTpsi}) is called the Darboux transformation.

For any eigenvalue, except the seed eigenvalue $\eta$, i.e. for $\lambda\not=\eta$, the transform (\ref{DTpsi}) of the two dimensional eigenspace of $H$ is also two-dimensional. However, when $\lambda=\eta$, operator $A$ annihilates the seed eigenfunction $\psi(z;\eta)=\varkappa$ and (\ref{DTpsi}) maps two dimensional eigenspace of $H$ into one-dimentional sub-eigenspace of $\hat H$.

Since, according to (\ref{AstarA}), $A^*$ annihilates $1/\varkappa$, and, by (\ref{DTH}), $\hat H=AA^*+\eta$, we see that $1/\varkappa$ is an eigenfunction of $\hat H$ corresponding to the eigenvalue $\eta$, i.e $\hat H[1/\varkappa]=\eta/\varkappa$. Then, the general solution $\hat \varkappa$ of the second-order differential equation $\hat H\hat\varkappa=\eta\hat\varkappa$ can be obtained from its particular solution $1/\varkappa$  by elementary methods. It equals 
\begin{equation}
\hat \varkappa = \frac{C}{\varkappa}\int \varkappa^2 dz, 
\label{DT0}
\end{equation}
where $C$ is an arbitrary constant, and the primitive of $\varkappa^2$ includes another arbitrary constant of integration. This extends the Darboux transformation (\ref{DTpsi}) of an eigenfunction to the case $\lambda=\eta$. Iterations of transformation (\ref{DT0}) are typically considered for $\eta=0$ and are called the Darboux transformations at zero energy level.

Darboux transformations can be iterated. Resulting transforms depend on seed eigenvalues $\eta=\eta_1, \eta=\eta_2, \dots $ chosen at different iterations. The Adler-Moser polynomials can be generated by such iterations, where the same seed eigenvalue $\eta_i=0$ is taken at each step. Indeed, rewriting the Tkachenko equation (\ref{Tkachenko}) in the Schrodinger form
$$
\left(-\partial^2-2(\log q)''\right)\left[\frac{p}{q}\right] =0 ,
$$
we can apply zero-enegy level Darboux transformation (\ref{DTH}, \ref{DT0}) to $\varkappa=p/q$ and $u=-2(\log q)''$ (see Eqs. (\ref{Schrodinger}, \ref{Heta}) for $\eta=0$). In this way we construct the sequence $\varkappa_n=P_{n+1}/P_n$, $u_n=-2(\log P_n)''$ starting from $\varkappa_0=z$ and $u_0=0$ (free Schrodinger operator $H_0=-\partial^2$). Indeed, rewriting (\ref{DT0}) in differential form with
\begin{equation}
\varkappa=P_{n}/P_{n-1}, \quad \hat\varkappa=P_{n+1}/P_n
\label{kappaP}
\end{equation}
we get the first-order recurrence relation for the Adler-Moser polynomials (\ref{AM3}). Therefore, given the initial conditions $P_0=1$, $P_1=z$, the Darboux transformation (\ref{DT0},\ref{kappaP}) generates all possible equilibrium configurations of charges with $Q_i\in\{-1,1\}$.

\section{$\Lambda=2$ Case and Third-Order Operators}
\label{ThirdOrder}

First, we briefly outline approach by Burchnal and Chaundy \cite{BC} adopted to the $\Lambda=2$ case in \cite{Lv}: Bilinear equation (\ref{L2}) with $q=q_n$ can be considered as a second-order linear equation for $p$ whose two linearly independent solutions are $p=p_{n-1}$ and $p=p_n$. Analogously, (\ref{L2}) is a linear equation for $q$, with coefficients defined by $p=p_n$, whose two linearly independent solutions are $q=q_n$ and $q=q_{n+1}$.  First-order recurrence relations (\ref{Bartman}) are the Abel identities relating linearly independent solutions. These identities, together with (\ref{L2}), imply that if the first linearly independent solution is a polynomial without multiple roots, so is the second solution (for detail see e.g. \cite{Lv, LY}). Therefore, from a polynomial solution $q_n, p_{n-1}$ of (\ref{L2}) we get a polynomial $p_n$. At the next step we get polynomial $q_{n+1}$ from solution $q_n,p_n$ etc. In this way, a sequence of solutions of (\ref{L2}) can be generated.

\begin{figure}
\centering
\includegraphics[width=155mm]{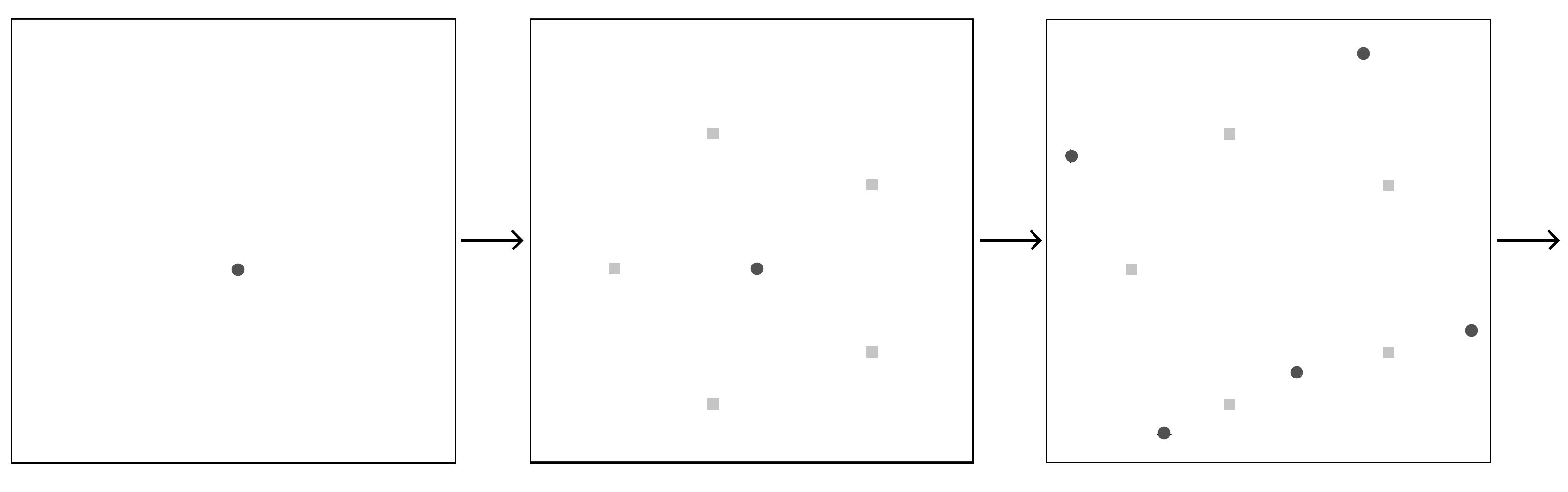}
\caption{$\Lambda=2$ case: Configurations corresponding to subsequent pairs $(q_1,p_0)$, $(q_1,p_1)$ and $(q_2,p_1)$. Roots of the first polynomial in the pair (positions of charges +2) are shown by the black circles, while roots of the second polynomial in the pair (positions of charges -1) are shown by gray squares. In this example $t_1=1, s_2=1+5{\rm i}$.}
\label{L2Roots}
\end{figure}

Note, that in difference from the $\Lambda=1$ case (where $n\in\mathbb{Z}_{\ge 0}$), in  the $\Lambda=2$ case the above procedure can be continued indefinitely for both increasing and decreasing $n$ ($n\in \mathbb{Z}$). Pairs $q_n,p_{n-1}$ and $q_n,p_n$ constitute complete set of polynomial solutions of (\ref{L2}) that do not have multiple/common roots \cite{Lv, LY}. Degrees of $q_n$ and $p_n$ are $n(3n-1)/2$ and $n(3n+2) $ respectively.
Examples of several first polynomials for $n\ge 0$ are
\begin{equation}
\begin{array}{ll}
q_0=1 & \quad p_0=1\\
q_1=z & \quad p_1=z^5+t_1\\
q_2=z^5+s_2z-4t_1 &
\begin{array}{r}
\;\; p_2={z}^{16}+{\frac {44}{7}}s_{{2}}{z}^{12}-32t_{{1}}{z}^{11}+22{s_{{2}}}^{2}{z}^{8}
-{\frac {2112}{7}}t_{{1}}s_{{2}}{z}^{7}\\+
1408{t_{{1}}}^{2}{z}^{6}+t_{{2}}{z}^{5}
-44{s_{{2}}}^{3}{z}^{4}+352t_{{1}}{s_{{2}}}^{2}{z}^{3}\\-1408s_{{2}}{t_{{1}}}^{2}{z}^{2}+
2816{t_{{1}}}^{3}z+t_{{2}}t_{{1}}-{\frac {11}{5}}{s_{{2}}}^{4}
\end{array}
\\
\dots & \quad \dots
\end{array}
\label{pqplus}
\end{equation}
etc, while for $n\le 0$
\begin{equation}
\begin{array}{lll}
q_0=1 & \quad & p_0=1 \\
q_{-1}=z^2+s_{-1} & \quad &  p_{-1}=z \\
\begin{aligned}
q_{-2}=z^7+7 s_{-1}z^5+35 s_{-1}^2z^3+ s_{-2}z^2\\
-35 s_{-1}^3z+ s_{-1} s_{-2}-\frac{5}{2}t_{-2}
\end{aligned}
& \quad &
\begin{aligned}
p_{-2}=z^8+\frac{28}{5}s_{-1}z^6+14s_{-1}^2z^4\\
+28 s_{-1}^3z^2+t_{-2}z-7 s_{-1}^4
\end{aligned}\\
\dots & \quad & \dots
\end{array}
\label{pqminus}
\end{equation}
etc, where $t_i$ and $s_i$ stand for arbitrary complex constants \footnote{We set constant corresponding to shift of $z$ to zero, so, without loss of generality, $q_1=p_{-1}=z$.} emerging in course of integrations of Abel identities. Examples of several first equilibrium configurations of the $n\ge 0$ sequence are shown on Figure \ref{L2Roots}.

Now we are ready to look for Darboux transformation in $\Lambda=2$ case: Similarly to the $\Lambda=1$ case, the bilinear equation (\ref{L2}), corresponding to $\Lambda=2$ case, can be rewritten in the Schrodinger form
\begin{equation}
\left(-\partial^2-6(\log q)''\right)\left[\frac{p}{q^2}\right]=0 ,
\label{SchrodingerL2}
\end{equation}
i.e. for sequences of $p_n$ and $q_n$ we have
\begin{equation}
\left(-\partial^2+u_n\right)\phi=0, \quad u_n=-6(\log q_n)'' ,
\label{Schrodinger2}
\end{equation}
where
\begin{equation}
\phi=p_n/q_n^2 \quad {\rm or} \quad \phi=p_{n-1}/q_n^2.
\label{varkappa2}
\end{equation}
However, sequence of the above second-order operators cannot be generated by Darboux transformations (\ref{DTH}), because now $u_{n\pm 1}\not=u_n-2(\log\phi)''$. To proceed, we note that according to the Abel identities (\ref{Bartman}), $p_n$ in (\ref{Schrodinger2}, \ref{varkappa2}) can be replaced with $q_{n+1}'q_n-q_{n+1}q_n'$ and we get the third-order differential equation
\begin{equation}
L\varkappa=0, \quad L=\partial^3-u_n\partial
\label{Ln}
\end{equation}
for $\varkappa$, where $\varkappa$ is any linear combination of $q_{n+1}/q_n$,  $q_{n-1}/q_n$ and $1$. Because $q_{n-1}, q_n, q_{n+1}$ have distinct degrees, this combination is the general solution of (\ref{Ln}). Since the linear combination $C_1q_{n+1}+C_0q_n+C_{-1}q_{n-1}$ is a polynomial, any solution of (\ref{Ln}) corresponds to an equilibrium configuration. Polynomial $C_1q_{n+1}+C_0q_n+C_{-1}q_{n-1}$ is a re-parametrized $q_{n+1}$ for $n>0$ or re-parametrized $q_{n-1}$ for $n\le 0$ respectively.

Thus, $\varkappa=q_{n\pm 1}/q_n$, where $n\ge 0$ or $n\le 0$ respectively, are general solutions of (\ref{Ln}). They correspond to complete set of equilibrium configurations of two species of charges with $Q_i\in\{-1,2\}$.

\section{Darboux Transformations for Third-Order Operators}

Darboux transformations for the third-order operators of the type (\ref{Ln}), i.e. transformations $L\to \hat L$, $\psi\to\hat\psi$ with the seed function $\varkappa$, where
\begin{equation}
L\psi=\lambda\psi , \quad L\varkappa=\eta\varkappa ,\quad L=\partial^3-u\partial , \quad \hat L\hat\psi=\lambda\hat\psi , \quad \hat L=\partial^3-\hat u\partial ,
\label{DT3}
\end{equation}
were found by Aiyer et al \cite{ABO}. Their derivation, via factorization, was presented by Athorne and Nimmo in \cite{AtNi}. Similarly to the $\Lambda=1$ case, we need ``zero-level" transformations (for $\lambda=\eta=0$) which we will find in this section.

Following \cite{AtNi}, we first review the case of $\eta=0$ and arbitrary $\lambda$. Operator $L$ can be presented through the product of the second and the first-order factors
\begin{equation}
L=BA,
\label{LBA}
\end{equation}
such that
\begin{equation}
B=\partial^2+f\partial-f'-\frac{f''}{f}, \quad A=\partial-f, \quad f=\frac{\varkappa'}{\varkappa} .
\label{BA}
\end{equation}
In contrast to the case of the Schrodinger operator, permutation of factors in (\ref{LBA}) maps $L$ into $\hat L$ of the similar type -- that is, an operator without the zeroth-order term, as in (\ref{DT3}) -- only if\footnote{This corresponds to $u={a}^{2} \left( 6\,c\,{\rm cn} \left( az+b,c \right) {\rm dn}\left( az+b,c \right)+6\,{c}^{2} \, {\rm sn}^2 \left( az+b,c \right) -{c}^{2} - 1 \right)$ and $\hat u=a^2( 6\,{c}^{2} {\rm sn}^2 \left( az+b,c\right)-{c}^{2}-1)$. In particular, in the rational limit, $u=0$ or $u=12/z^2$, where $v=2/z$ or $v=-2/z$ respectively.} $v=2ac\,{\rm sn}(az+b,c)$, where $a,b$ and $c$ are arbitrary constants. In general, to obtain an operator of the same type, several permutations and re-factorizations are required (see \cite{LY} for more details). Instead of studying these intermediate operations separately, we consider the transformation
\begin{equation}
A\psi=\hat A\hat\psi, \quad \hat A=\partial-\hat f .
\label{AphAhp}
\end{equation}
Multiplying equation $L\psi=\lambda\psi$ by $A$ on the left and taking (\ref{LBA}) and (\ref{AphAhp}) into account we get
$$
AB\hat A\hat\psi=\lambda \hat A\hat\psi . 
$$
Demanding that
\begin{equation}
AB=\hat A \hat B , \quad \hat B=\partial^2+\hat f\partial-\hat f'-\frac{\hat f''}{\hat f}
\label{ABhAhB}
\end{equation}
we see that, up to element of kernel of $\hat A$,
$$
\hat B \hat A\hat\psi=\lambda \hat \psi . 
$$
Equation (\ref{ABhAhB}) has a solution $\hat f = -f$ and the Darboux-type transform $\hat L=\hat B\hat A$ is obtained from $L=BA$ by the involution\footnote{This is similar to the Schrodinger operator case where the Darboux transformation is the involution $A=\partial-f \leftrightarrow A^*=-(\partial+f)$, cf (\ref{AstarA}).} $f\to -f$. From (\ref{LBA}, \ref{BA}) it follows that for such a transformation
\begin{equation}
\hat u = u - 6 (\log \varkappa)'' .
\label{DTu}
\end{equation}
Since $\hat A=\partial+f=\partial+\varkappa'/\varkappa$, function $\varkappa^{-1}$ is an eigenfunction of $\hat L=\hat B\hat A$ corresponding to $\lambda=0$:
$$
\hat L \varkappa^{-1}=0 .
$$
Similarly to the Schrodinger case, the complete zero-level transform $\hat\varkappa$ of $\varkappa$ is a general solution of equation
\begin{equation}
\hat L\hat\varkappa=0.
\label{hLhkappa}
\end{equation}
It can be obtained from its particular solution $\varkappa^{-1}$ by elementary methods: Fist, we rewrite (\ref{hLhkappa}) as
$$
\hat \phi''-\hat u \hat \phi=0, \quad \hat\phi=\hat\varkappa' .
$$
One of the linearly independent solution of the above second order equation equals $(\varkappa^{-1})'$. Then, its general solution is
\begin{equation}
\hat\phi = C(\varkappa^{-1})'\int \frac{dz}{\left((\varkappa^{-1})'\right)^2} ,
\label{hatphi}
\end{equation}
where $C$ is an arbitrary constant and the primitive includes an arbitrary constant of integration. To get $\hat\varkappa$, we have to integrate (\ref{hatphi}), i.e.
$\hat\varkappa=\int\hat\phi dz=C\int \left(\int \frac{dz}{\left((\varkappa^{-1})'\right)^2}\right)d\frac{1}{\varkappa}$. Integrating by parts, we obtain, up to multiplication by an arbitrary constant,
\begin{equation}
\hat\varkappa = \int \frac{\varkappa^3}{(\varkappa')^2}dz - \frac{1}{\varkappa}\int \frac{\varkappa^4}{(\varkappa')^2}dz,
\label{DTkappa}
\end{equation}
where the primitives include arbitrary constants of integration. Thus, we have found the zero-level Darboux transformations (\ref{DTu}, \ref{DTkappa}) for the third-order operators of the type $\partial^3-u\partial$.

\section{$\Lambda=2$ Configurations and Darboux Transformations}
\label{DTEq}

As was shown in section \ref{ThirdOrder}, the general solution\footnote{Here, for simplicity, we consider situation with increasing $n\ge 0$.} of equation $ L\varkappa=0$, where $L=\partial^3+6(\log q_{n-1})''\partial$, is $\varkappa=q_n/q_{n-1}$. According to (\ref{DTu}), the Darboux transform of $L$ equals $\hat L=\partial^3+6(\log q_n)''\partial$. Since the transform $\hat\varkappa$ of $\varkappa$ is the general solution of $\hat L\hat\varkappa=0$, we conclude that $\hat\varkappa=q_{n+1}/q_n$ and the Darboux transformations (\ref{DTkappa}) generate sequence $q_n$.

Note that the more general fact holds: The first-order recurrence relations (\ref{Bartman}) are, in fact, the Darboux transformations (\ref{DTkappa}) rewritten in differential form. Derivation of these relations doesn't need to involve bilinear equation (\ref{L2}). 

Indeed, let us write
\begin{equation}
\hat\varkappa=q_+/q, \quad \varkappa=q/q_- ,
\label{qpmq}
\end{equation}
where $q_+$, $q$ and $q_-$ are some functions of $z$. Then
\begin{equation}
\hat\varkappa'=p/q^2, \quad \varkappa'=p_-/q_-^2 ,
\label{kprime}
\end{equation}
where
\begin{equation}
p = q_+'  q-q_+ q', \quad p_- = q'  q_- - q q_-' .
\label{Abel1}
\end{equation}
Substituting (\ref{qpmq},\ref{kprime}) into transformation (\ref{DTkappa}), or equivalently, substituting $\hat \phi=p/q^2$, $(\varkappa^{-1})'=-\varkappa^{-2} p_-/q_-^2$ and $\varkappa=q/q_-$ into (\ref{hatphi}) and rewriting result in the differential form, we obtain
\begin{equation}
q^4 \propto p'  p_- - p p_-' .
\label{Abel2}
\end{equation}
Equations (\ref{Abel1}, \ref{Abel2}) are, modulo nonessential constant factors, the first-order recurrence relations (\ref{Bartman}) with $q_+=q_{n+1}$, $q=q_n$, $q_-=q_{n-1}$ and $p=p_n$, $p_-=p_{n-1}$. Together with initial conditions $q_0=p_0=1$, they produce solutions of (\ref{L2}). There also exist other initial conditions for which solutions of (\ref{Bartman}) are polynomials that are not solutions of (\ref{L2}). They will be considered in the next section.

In summary: All polynomial solutions of the bilinear equation (\ref{L2}) that do not have common/multiple roots can be generated by iterations of transformation (\ref{DTkappa}) with
\begin{equation}
\hat \varkappa = q_{n+1}/q_n, \quad \varkappa= q_n/q_{n-1}
\label{iterations}
\end{equation}
together with equation (cf. (\ref{Abel1}))
\begin{equation}
p_n \propto q'_{n+1}q_n-q_{n+1}q_n' 
\label{Abel}
\end{equation}
and initial conditions $q_0=p_0=1$.  Thus, starting from $\varkappa=\varkappa_0=z$, iterations of transformation (\ref{DTkappa}) produce rational functions $\varkappa_n=q_{n+1}/q_n$, such that zeros of $\varkappa'_n$ correspond to positions of charges with $Q_i=-1$, while poles  of $\varkappa'_n$ correspond to positions of charges with $Q_i=2$.

\section{Darboux Transformations and Terminating Configurations}
\label{Even}

Introducing factorizable function
\begin{equation}
\phi={\rm const}\prod_{i=1}^N(z-z_i)^{Q_i},
\label{psiQ}
\end{equation}
one can easily show that the equilibrium conditions (\ref{equilibrium}) are equivalent to the absence of simple poles in the potential $u$ of the Schrodinger equation $(-\partial^2+u)\phi=0$.

In the case of systems related to the second-order operators, from (\ref{DTH}) it follows that the transform $\hat u$ of $u$, where $\varkappa=\phi$, do not have simple poles. Therefore, provided the transform $\hat\varkappa$ of $\varkappa$ is also factorizable, it corresponds to an equilibrium configuration.

Similar fact takes place in the third-order case. Indeed, due to (\ref{DTu}), simple poles are absent in the potential of the Schrodinger equation $(-\partial^2+\hat u)\hat\phi=0$, $\hat \phi=\hat\varkappa'$. Therefore, Darboux transformation maps equilibrium configuration into another equilibrium configuration in both second and third-order cases, provided the transform $\hat\varkappa$ of $\varkappa$ is factorizable.

The set of equilibrium configurations that can be generated by Darboux transformations is bigger than the set of solutions (\ref{Tkachenko}) or (\ref{L2}) and, apart from configurations of two species of charges, also comprises configurations with three species: Factorizable $\varkappa$ need not to necessarily correspond to potentials $u=-2(\log q)''$ in the second-order or $u=-6(\log q)''$ in the third order case.

As was mentioned in section \ref{DT2}, the first-order recurrence relation (\ref{AM3}) for $P_n$ follows from the Darboux transformation (\ref{DT0}, \ref{kappaP}). Bilinear equation (\ref{Tkachenko}) is not involved in this derivation. Therefore, to get a sequence of equilibrium configurations, the initial conditions for (\ref{AM3}) need not necessarily to be those for the Adler-Moser polynomials $P_0=1$, $P_1=z$.  A family of sequences with other initial conditions was introduced in work by Duistermmat and Grunbaum \cite{DG} in the context of bi-spectral problem. In relation to vortex equilibria this family appeared in work by O’Neil and Cox-Steib \cite{ONCS} (also see \cite{KWCC}). There, sequences start from $P_0=1$, $P_1=z^{1/2}$. In difference from the Adler-Moser case, these sequences terminate as a non-factorizable $\varkappa$ emerges at some step of recurrence. Examples of first several sequences are:
\begin{equation}
\begin{array}{l}
P_0=1, \quad P_1=z^{1/2}, \quad  P_2=z^2+t_1 . \\
P_0=1, \quad P_1=z^{1/2}, \quad  P_2=z^2, \quad P_3=z^{9/2}+t_2z^{1/2}, \quad P_4=z^8+6t_2z^4+t_3z^2-3t_2^2 . \\
\dots
\end{array}
\label{Peven}
\end{equation}
etc. Here, a half-integer charge of the third specie is present at the origin $z=0$.

Similar situation occurs in the case of the third-order operators. As we saw in section \ref{DTEq}, the Darboux transformations (\ref{DTkappa}, \ref{iterations}), together with (\ref{Abel}), result in the first-order recurrence relations (\ref{Bartman}). For two species of charges initial conditions are $p_0=q_0=1$. However, there exist other initial conditions, where terminating sequences of equilibrium configurations, that are not solutions of (\ref{L2}), are generated by the Darboux transformations. Such sequences of solutions were found in \cite{ONCS} (also see \cite{DK1}, \cite{KWCC}). For example, for $q_0=1$, $p_0=z^2$ the first several sequences are
$$
\begin{array}{l}
q_0=1, \quad p_0=z^2, \quad q_1=z^3+t_1 . \\
q_0=1, \quad p_0=z^2, \quad q_1=z^3, \quad p_1=z^2(z^9+t_2), \quad q_2=z^9+t_3z^3-2t_2 .\\
\dots
\end{array}
$$
and so on.

\section{Conclusions and Open Problems}

In this article we found Darboux-type transformations that generate complete set of equilibrium configuration of two species of vortices with circulation ratio $-2$. They also generate sets of configurations, where an additional charge of the third specie is present.

In conclusion, we would like to mention open problems and outline several generalizations of systems studied in the present article.

First, we would like to mention problem of determinant representation of functions corresponding to the equilibrium configuration. Wronskian representation of the Adler-Moser polynomials is known since work by Burchnall and Chaundy \cite{BC}. This representation can be considered as a zero-level degeneration of the Crum theorem for sequences of the Darboux transformations (for details see e.g. \cite{LY, MS}). Using such an approach (see e.g. \cite{LY}) one can also derive Wronskian representation for terminating configurations of type (\ref{Peven}).

In the third-order case, one has to deal with Pfaffians, rather than Wronskians \cite{KVdL, Ni}. ``Pfaffian analog" of the Crum theorem exists for a sequence of Darboux transformations associated with the set of distinct seed eigenvalues \cite{Ni}. However, it is unclear how to deal with its zero-level degeneration, because, in difference from the second-order case, the eigenfunctions must satisfy certain boundary conditions \cite{ABO, AtNi, Ni}.

Recently, a Pfaffian representation of polynomial $\tau$-functions of the Sawada-Kotera hierarchy was found  by Kac and Van de Leur in \cite{KVdL} via free-fermion approach  (for more details on Sawada-Kotera and Kaup-Kupershmidt hierarchies, see e.g. \cite{ABO, FG, KVdLKK, KVdL} and references therein).

It is known from the theory of integrable hierarches that, for a rational Lax operator $L$, there exists an eigenfunction $\varkappa=\theta/\tau$ corresponding to zero eigenvalue, where $\tau$ is a polynomial $\tau$-function and $\theta$ is also a polynomial. In the case of the Sawada-Kotera hierarchy $L=\partial^3-u\partial$, where $u=-6(\log\tau)''$ is a solution of the hierarchy (see e.g. \cite{KVdL} and references therein).  Equation $L\varkappa=0$, rewritten in the Schrodinger form, is equivalent to the bilinear equation (\ref{L2}) for $q=\tau$ and $p=\tau'\theta-\tau\theta'$. Then, since $p$ and $q$ are polynomials, polynomial $\tau$-functions of the Sawada-Kotera hierarchy correspond to equilibrium configurations.

Therefore, the result by Kac and Van de Leur provides a Pfaffian representation for equilibrium configurations of two species of charges with $Q_i\in\{-1,2\}$.

It was shown by Fordy and Gibbons \cite{FG} that solutions $u$ of the Sawada-Kotera and solutions $v$ of the Kaup-Kupershmidt hierarchy\footnote{The Lax operator of the Kaup-Kupersmidt hierarchy equals $L=\partial^3+v\partial+\frac{1}{2}v'$ .} are related by the Miura transormation
$$
u=w'+w^2, \quad v=2w'-w^2.
$$
It easy to see that for rational solutions of the Sawada-Kotera hierarchy $w=\phi'/\phi$, $\phi=p/q^2$. Then, taking (\ref{L2}) into account, we conclude that the Miura transform of $u$ equals $v=3(\log p)''$. The last equation is nothing but formula for $\tau$-function of the Kaup-Kupershmidt hierarchy, where $\tau=p$. Therefore, polynomial $\tau$-functions of the Kaup-Kupershmidt hierarchy correspond to $\Lambda=2$ equilibrium configurations.

Kac and Van de Leur found a determinant representation for polynomial $\tau$-functions of the Kaup-Kupershmidt hierarchy \cite{KVdLKK}. Thus, apart from the Pfaffian, $\Lambda=2$ equilibrium configurations also have determinant representation.

In this light, it would be interesting to explore possibility of existence of Pfaffian or determinant representation for terminating configurations presented in section \ref{Even} for the third-order case.

Finally, we would like to outline several generalizations of systems considered in this paper: One of generalizations extends (\ref{equilibrium}) to the case of configurations translating with constant velocity or equilibria of charges in homogeneous background electric field $k$:
$$
k+\sum_{j=1, j\not=i}^N\frac{Q_j}{z_i-z_j}=0, \quad i=1\dots N .
$$
Extension of the bilinear equation (\ref{bilinear}) for this case, rewritten in the Schrodinger form, is
$$
\left(\partial^2+\Lambda(\Lambda+1)(\log q)''\right)\Phi=k^2\Phi, \quad \Phi=\frac{p(z,k)}{q(z,k)^\Lambda}e^{kz} .
$$
When $\Lambda=1$, $\Phi(z,k)$ is the ``rational" Baker-Akhiezer function of the KdV hierarchy. It can be obtained by application of sequence of the $\eta=0$ Darboux transformations (\ref{DTpsi}, \ref{DT0}) to the eigenfunction $\psi=e^{kz}$ of the free Schrodinger operator corresponding to the eigenvalue $\lambda=k^2$. In this way one gets all possible translating configurations for $Q_i\in\{-1,1\}$  (see e.g. \cite{LY}).

In contrast to the $\Lambda=1$ case, the Darboux transformation method does not work when $\Lambda=2$ and $k\not=0$. This is because the $\eta=0$ Darboux transformations of $e^{kz}$ are now associated with another ``rational" Baker-Akhiezer function $\Psi(z,k)$, a such that
$$
L\Psi=k^3\Psi,
$$
where $L=\left(\partial^2+6(\log q)''\right)\partial$ is of the third order. Only when $k=0$, the above eigenvalue problem reduces to solution of the Schrodinger equation (\ref{SchrodingerL2}), which is equivalent of (\ref{L2}). Therefore, here Darboux method works only for non-translating (i.e. $k=0$) configurations. Thus, question of classification of translating configurations remains open for $\Lambda=2$.

Further generalization deals with the periodic configurations (``vortex streets") satisfying
$$
k+\sum_{j=1, j\not=i}^N Q_j\cot(z_i-z_j)=0, \quad i=1\dots N .
$$
When $\Lambda=1$, this generalization is associated with the soliton solutions of the KdV hierarchy and ``trigonometric" Baker-Akhieser function $\Phi=p(z,k)/q(z)e^{kz}$, where now $q$ and $p$ are trigonometric polynomials in $z$ (\cite{ HV, KC, Lbhe}, for review see e.g. \cite{LY}). Trigonometric Baker-Akhiezer function is obtained from the eigenfunction $e^{kz}$ (with eigenvalue $\lambda=k^2$) of the free Schrodinger operator $\partial^2$ by iterations of the Darboux transformations: First with $\eta=k_1^2$, then transformation with $\eta=k_2^2$, etc. Here $k_i$ are integers, such that $0<k_1<k_2<\dots $.

When $\Lambda=2$, translating vortex streets configurations cannot be generated by Darboux transformations: Even in the non-translating case $k=0$, only short terminating sequences are found \cite{LY}. This is because the operator $L$ is now of the third order. Moreover the ``free" operator $L_0=-\partial^3$ does not have trigonometric eigenfunctions and one cannot expect to find non-terminating sequences here. It would be interesting to explore the $\Lambda=2$ vortex street configurations using alternative methods.


\end{document}